\documentclass[aps,prl,twocolumn,showpacs,groupedaddress]{revtex4}
\usepackage{graphicx}
\usepackage{dcolumn}
\usepackage{bm}
\usepackage{amssymb}
\usepackage{color}
\begin{document}
\title{Generation of $1/f$ noise motivated by a model for musical
melodies}

\author{Martin Grant and Niloufar Faghihi}

\date{\today}

\affiliation{Physics Department, Rutherford Building, 3600 rue University,
McGill University, Montr\'eal, Qu\'ebec, Canada H3A 2T8}

\begin{abstract}

We present a model to generate power spectrum noise with intensity
proportional to $1/f$ as a function of frequency $f$.  The model
arises from a broken-symmetry variable which corresponds to absolute
pitch, where fluctuations occur in an attempt to restore that
symmetry, influenced by interactions in the creation of musical
melodies.

\end{abstract}

\maketitle

\section{Introduction}

Correlations of melodies in music follow a power spectrum where the
intensity of that spectrum varies in inverse proportion to the
frequency $f$, called $1/f$ noise \cite{manaris-05}.  This
characteristic spectrum is followed by musical rhythms
\cite{levitin-12}, pitch \cite{manaris-05,lu-2013}, and consonance
fluctuations in music \cite{levitin-15}.  The power spectrum is
related to autocorrelation functions \cite{Voss-75,Gardner}, and hence
the $1/f$ power spectrum signifies a correlation in, for example, the
fluctuating pitch on a time scale over which the power spectrum is
$1/f$ \cite{voss-78}.  The $1/f$ power spectrum in music produces a
balance between the predictable sound intensity of a random walk
$1/f^2$ and random white noise $1/f^0$.  This balance between the
expected and surprise is said to be psychologically pleasing for the
audience \cite{levitin-12,levitin-10, Levitin_Book}.

In this paper we show that this noise spectrum can be recovered if,
firstly, we consider pitch as a broken-symmetry variable, which we
believe to be a necessary feature of a mathematical theory of musical
melodies, and if, secondly, we include interactions between musicians
\cite{gilden-95,hennig-12,hennig-14,ruiz-14}, which we find to be a
sufficient feature.

An understanding of music is rooted in the way the brain works
\cite{wu-09}, and is different in kind from most other arts.  A useful
contrast is to painting where, often, a representation of the outside
world is seen, over-colored in red, yellow, and blue due to our eyes'
particular sensitivity to those colors.  Of course, a painting is not
a random snapshot of what is ``outside'' in the world, there is an
emotional and intellectual resonance ``inside'' the artist's mind from
which comes the choice of subject or the manner of presentation
\cite{Shepard-94}.  Therein lies the art, but its root is
representational. The same could be said of plays, sculpture, and
indeed of most arts: the artifact would be some representation of what
is outside --- exaggerated or highlighted somehow --- which has an
emotional and intellectual resonance inside; these arts have an
external referent.

Contrast this to music.  Imagine watching a film of a jazz quartet,
with the sound turned off.  Further, imagine that the film shows the
score of the piece.  As a musician begins playing, and one sees the
score, the performance is not that different from watching a
mathematician at a blackboard, flawlessly writing complex equations.
As the other musicians begin, it would be as if our mathematician is
joined by three others, each writing varied but related theorems at
other blackboards, all in synchrony.  This is strange, but not the
strangest part.  What is bizarre is the audience, perhaps humming
along or tapping their fingers in time as they themselves interact
with the musicians \cite{repp-05} and even predicting the 
onset times of upcoming musical events \cite{rankin-09,rankin-14}.
As the equations multiplied on the blackboards,
one would notice that the people listening to the four mathematicians
would experience intense feelings of joy, followed by feelings of
loss, then elation again, and so on, as they followed the tracings of
algebra.

As this thought experiment shows, the representational part of music,
based on what is ``outside'' in the world, is minimal.  Musical
performances are not snapshots of a natural soundscape.  Listeners'
intense reaction to, and involvement with music shows the performance
is ``inside'' the audience's minds.  The snapshot taken --- however
exaggerated or highlighted with emotional and intellectual resonance
--- is a snapshot of their brains.  That is music, and the direct
route it has to our brains.

To understand music, then, is to understand some aspects of the
neuroscience of the brain \cite{large-08,large-09,palmer-97};
music provides a concrete way of addressing
how, for example, decisions are made in the context of the creation of
musical melodies.  Furthermore, we note that music is particularly
well suited to a mathematical analysis as, unlike other art forms, it
is straightforwardly quantifiable in terms of time and pitch
\cite{campbell-86, schroeder-02}.  As mentioned above, it is well
known that musical melodies obey $1/f$ noise.  However there is no
theory giving rise to this, and demonstrations of artificial music
derived from a $1/f$ power spectrum make use of, for example, filtered
white noise \cite{Gardner} for which no potential neurological
significance can be attached.  We also note that the phenomena of
$1/f$ noise, although well documented, is in seeming contradiction to
the notions of time signatures, key signatures, and the different
modes and scales in different cultures.  We will not address that in
this paper, except in the sense of attempting to find the simplest
model consistent with musical melodies which gives rise to $1/f$
noise; further, and crucially, we make use of the fact that a musical
melody can be recognized without regard to the key of that melody,
demonstrating the imperfect detection of absolute pitch.  We will
obtain a $1/f$ spectrum which is naturally motivated by musical
melodies: the two essential ingredients of our model are the
arbitrariness of absolute pitch, which is a necessary feature of a
treatment of music, and interactions, which we find to be a sufficient
feature.

\section{Broken-symmetry variables}

In this section, we will review the properties of broken-symmetry
variables \cite{Forster,KPZ,Grossmann,DDS,Bak,Bak_book}.  When, for
example, a free energy has a continuously broken symmetry, this gives
rise to a broken-symmetry variable $h$, whose fluctuations act to
restore that symmetry.  The modes of a broken-symmetry variable have
no energy gap to begin propagating, and hence the correlations of $h$
are power-laws, without scale.  In particular, in Fourier space, as a
function of wavenumber $q$, it can be shown that a broken-symmetry
variable, often called a Goldstone variable, satisfies $\langle |\hat
h(q) |^2 \rangle \propto 1/q^2$, for small $q$, where $\hat h$ is the
Fourier transform of $h$, defined below, and the brackets denote an
average.

We will now give a description of the broken-symmetry
variable associated with surfaces of coexisting phases, and then adapt
it to musical melodies.  For surfaces, the broken symmetry is
translational invariance in the direction normal to the interface, the
broken-symmetry variable is the local height of the interface, and the
dynamical modes are called capillary waves or ripplons.  Consider a
surface in the $d$-dimensional ${\vec x}$ plane of a
$(d+1)$-dimensional system ${\vec r} = ({\vec x}, y)$.  The position
of the surface, assuming small fluctuations and therefore no droplets
or overhangs, is $y=h({\vec x})$, where $h$ is the local height of the
surface.  The probability of being in a state $h({\vec x})$ is
proportional to $e^{-F(h)}$, where $F$ is the free energy in units of
the temperature.  If the free energy does not depend on the absolute
position of the interface, which corresponds to translational
invariance of space in the direction normal to the interface, it is
invariant under $h \rightarrow h +{\rm const}$.  Hence it can only
depend on $h$ through gradients.  To lowest order in gradients, we
have 
\begin{equation}
F \propto \int d \vec x (\nabla h)^2.
\end{equation}
Note that the free energy is minimized by minimizing variations in
$h$, any constant $h$ minimizes $F$.  This form, being quadratic, can
be diagonalized and solved as a product of independent Gaussians by
using Fourier transforms: 
\begin{equation}
\hat h(\vec q) = \int d\vec x e^{-i\vec q\cdot \vec x} h(\vec x)
\end{equation}
where we can write
\begin{equation}
h(\vec x) = \int \frac{d\vec q}{(2\pi)^d}  
	e^{i\vec q\cdot \vec x} \hat h(\vec q),
\end{equation}
since the Dirac delta function ensures completeness, 
\begin{equation}
\delta (\vec x- \vec x') = 
   \int  \frac{d\vec q}{(2\pi)^d}  e^{i\vec q\cdot (\vec x - \vec x')}.
\end{equation}
The solution is 
\begin{equation}
\langle h(\vec q) h(\vec q') \rangle = (2\pi)^d \delta(q+q') \hat G(q),
\end{equation}
where 
\begin{equation}
\hat G(q) \propto \frac{1}{q^2}.
\end{equation}
This is the standard result for a broken-symmetry Goldstone variable
\cite{Forster}.  As the integrals are Gaussian, all moments can be
obtained from the second moment.  In real space, correlation functions
satisfy 
\begin{equation}
\langle h(\vec x)h(0)\rangle = \int \frac{d\vec q}{(2\pi)^d} 
	e^{i\vec q \cdot \vec x} \hat G(q).
\end{equation}
Translational and rotational invariance of space in the plane {\it
parallel\/} to the interface gives rise to the delta function $\delta
(\vec q + \vec q')$, and implies that $\langle h(x)h(0)\rangle =
\langle h(\vec x + \vec x_0 ) h (\vec x_0)\rangle$, where $\vec x_0$
corresponds to the arbitrary origin, and can be ignored.

This general result can be obtained in other ways.  In
particular, and which is necessary for our purposes, the noise source
does not need to be thermal: the derivation of the Kardar-Parisi-Zhang
equation \cite{KPZ,Grossmann,DDS} provides an example.  If the rate in
change of time $\tau$ of a broken-symmetry variable has the invariance
$h \rightarrow h +{\rm const}$, and we endeavour to find the rate of
change of $h$ via a gradient expansion in $h$, and to lowest order in
time derivatives,  we obtain 
\begin{equation} 
\frac {\partial h}{\partial \tau } \propto \nabla^2 h + \mu, 
\end{equation} 
where $\mu$ is a noise.  By the central-limit theorem, the only
nontrivial moment of the noise satisfies: 
\begin{equation} 
\langle \mu(\vec x, \tau) \mu (\vec x', \tau') \rangle
	\propto  \delta (x -x') \delta (\tau - \tau').
\end{equation} 
Note that, for our purposes, we will not need to consider the
higher-order term in the Kardar-Parisi-Zhang equation which breaks the
symmetry $h \rightarrow -h$.  In steady state, one again recovers the
same probability and free energy, and the same results, including
$\hat G \propto 1/q^2$.

Finally, let us define a partial Fourier transform using $\vec q =
(q_x, \vec k)$, such that $q_x$ is one-dimensional, and $\vec k$ is
$(d-1)$-dimensional.  Namely, let 
\begin{equation}
P(x, \vec k) = \int \frac{d q_x}{2\pi} e^{iq_x \cdot x} \hat G(q_x, \vec k).
\end{equation}
Using $\hat G = 1/q^2 = 1/(q_x^2 + k^2)$ gives straightforwardly,
\begin{equation} 
P(x, \vec k) = \frac{e^{-|k|x}}{|k|},
\end{equation}
so that
\begin{equation} 
P(0, \vec k) = 1/{|k|}.
\end{equation}
Note that the $x=0$ axis is equivalent to any constant $x$ axis from
translational invariance in the plane parallel to the interface.

\section{Model of melody}

A necessary feature of a theory of music is the imperfect detection of
absolute pitch.  Unlike vision, where the wavelengths corresponding to
red, blue and yellow are privileged for most people, the absolute
frequencies of notes are arbitrary for most.  For example, a musical
melody is recognized without regard to the key in which it is
performed.  Having a specific pitch, then is the same as having a
specific value for a broken-symmetry variable.  Hence we identify
absolute pitch, in essence the key of a melody, as a broken-symmetry
variable, and so the highness $h$ of a pitch in a melody will behave
with the correlations of a broken-symmetry variable.

This is a necessary component to understanding musical melodies.  We
can imagine a solitary musician sitting by the piano, playing notes at
random, close to a previous pitch.  Following the nergy function
above, the decision on what note to follow a given note is given by
the constraint enforced by gradients of $h$ above, which favours
nearby notes; the concept of nearby is used in the most simple sense
here, and more refined descriptions exist \cite{Shepard}.  If the
highness of the pitch is also called $h$, the same notation as for the
example involving surfaces above, and the space in which the notes are
played is one-dimensional time $t$, we have 
\begin{equation} 
\langle h(t)h(0)\rangle = \int \frac{df}{(2\pi)}
	e^{i f \cdot t} \hat G(f),
\end{equation} 
where we have introduced the frequency $f$ replacing the Fourier
wavenumber $q$ above, and 
\begin{equation} 
\hat G(f) \propto \frac{1}{f^2}.  
\end{equation} This is the frequency spectrum of a random walk, not
$1/f$ noise.  Hence, recovering $1/f$ noise will require more than
the imperfect detection of pitch: we will introduce another element,
interactions, which will act to temper these variations in frequency.  

As such, let us insist that the solitary musician listens to other
musicians, all of whom are also attempting to play melodies in the
same manner as our first musician.  Further, we will assume the
musicians temper each others' choices of notes such that notes chosen
are constrained not only to be close to a previous note, but to be
close to those played by the other musicians.   Now the dimension of
space is two-dimensional and the frequency spectrum changes: one
dimension is time $t$ and the second dimension is the direction of
interaction with other musicians at a given time called $x$, where we
can conveniently measure the strength of the interactions in units of
$t$ and $x$.  This model provides  a minimal interaction that one
musician can have with others.  From above, we immediately obtain the
correlation function 
\begin{equation} 
P(x, f) = \frac{e^{-fx}}{f}.
\end{equation}
Hence, considering the notes on the $x=0$ axis corresponding to a
chosen musician, we have 
\begin{equation} 
P(0, f) = 1/f.
\end{equation}
This is our main result, recovering $1/f$ noise.  It arises from one
necessary condition, imperfect absolute pitch detection, and one
sufficient and we believe plausible condition, interactions, which has
been identified and noted in the literature
\cite{gilden-95,hennig-12,hennig-14,ruiz-14}.  There are two further
implications of our analysis which are testable.  Firstly, the
exponentially-damped correlations of Eq.\ (15) correspond to
correlations across nonzero $x$ and frequency.  Secondly, for $t=0$,
which corresponds to any fixed time, we find that  there are power-law
correlations in the Fourier transform of $x$, $q_x$, following
$1/|q_x|$.

\begin{figure}[ht]
\begin{center}
\includegraphics[width=8cm]{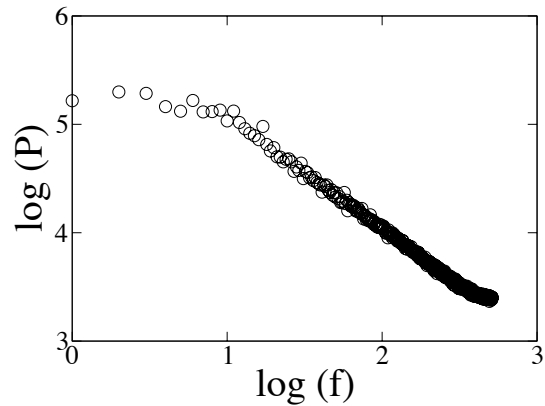}
\caption{
The power $P$ spectrum of the solid-on-solid
model showing $1/f$ noise as a function of $\log (P)$
versus $\log (f)$.
}
\label{fig:spectrum}
\end{center}
\end{figure} 

As an example of how readily a $1/f$ spectrum can be obtained, we show
the results of a Monte Carlo simulation of the equilibrated
solid-on-solid model \cite{Swendsen,Grossmann}, which is known to be a
good model of capillary wave fluctuations.  Each lattice site of a
two-dimensional square grid can have any integer height $h_i$, where
$i = 1, 2, \ldots N$ runs over all sites of the square lattice of $N$
sites.  The energy of interaction is proportional to $\Sigma_i |h_i
-h_{nn}|$, where $nn$ denotes the nearest neighbours of $i$.  In
Figure \ref{fig:spectrum} we show the power spectrum of the
equilibrated solid-on-solid model with $N = 1024\times1024$ at noise
intensity $T=4$, in units of the interaction constant.   This is
the averaged power spectrum of one ``row'' of the model, corresponding
to the time axis of the evolving melody.

In conclusion, we have presented a simple model which recovers $1/f$
noise for musical melodies from one necessary condition, imperfect
absolute pitch detection, and one plausible sufficient condition,
interactions.  A more refined theory of  musical melodies would be of
value, but this study provides a first step, and we believe the
features we have identified will play roles in any such refined
theories.  

\section{Acknowledgements}

We thank Dan Levitin, Charles Gale and Caroline Palmer
for useful discussions.  This
work was supported by the Natural Sciences and Engineering Research
Council of Canada and by {\it le Fonds de recherche du Qu\'ebec ---
Nature et technologies \/}.

\end{document}